# Optically defined cavities in driven-dissipative photonic lattices

O. Jamadi[1†*], B. Real[1*], K. Sawicki[2], C. Hainaut[1], A. González-Tudela[3], N. Pernet[4],

I. Sagnes[4], M. Morassi[4], A. Lemaître[4], L. Le Gratiet[4], A. Harouri[4], S. Ravets[4], J. Bloch[4]

and A. Amo[1‡]

[1]*Université de Lille, CNRS, UMR 8523 – PhLAM – Physique des Lasers, Atomes et Molécules, Lille, France*

[2]*Institute of Experimental Physics, Faculty of Physics, University of Warsaw, Pasteura St. 5, 02-093 Warsaw, Poland*

[3]*Institute of Fundamental Physics IFF-CSIC, Calle Serrano 113b, 28006 Madrid, Spain.*

[4]*Université Paris-Saclay, CNRS, Centre de Nanosciences et de Nanotechnologies, 91120, Palaiseau, France*

† *email :* omar.jamadi@gmail.com

‡ *email :* alberto.amo-garcia@univ-lille.fr

*\* These authors contributed equally*

**The engineering of localised modes in photonic structures is one of the main targets of modern photonics. An efficient strategy to design these modes is to use the interplay of constructive and destructive interference in periodic photonic lattices. This mechanism is at the origin of defect modes in photonic bandgaps, bound states in the continuum and compact localised states in flat bands. Here we show that in lattices of lossy resonators, the addition of external optical drives with controlled phase enlarges the possibilities of manipulating interference effects and allows designing novel types of localised modes. Using a honeycomb lattice of coupled micropillars resonantly driven with several laser spots at energies within its photonic bands we demonstrate the localisation of light in at-will geometries down to a single site. These localised modes can be seen as fully reconfigurable optical cavities with the potentiality of enhancing nonlinear effects and of controlling light-matter interactions with single site resolution.**

Engineering the localisation of light in dielectric structures is one of the main targets of modern micro- and nano-photonics. From a fundamental point of view, the study of localisation in photonic materials has allowed understanding the subtle interplay of disorder, periodicity and quasi-crystalline order [1–3]. Localized modes are useful to enhance light-matter interactions, to increase the non-linear response of a material and to store information in a reduced volume. A very successful strategy to engineer localised modes in optics is to use lattices of coupled waveguides, coupled resonators, or photonic crystals. In these periodic structures, localised modes at predefined wavelengths and spatial locations can be engineered by carefully designing the bulk eigenmodes or by introducing defects in bandgaps. Examples of the first type are the modes localised by disorder (Anderson localization) in a lattice [1,2], the compact localised states of a flat band [4–6], the bound-states in the continuum [7–10], and localised modes in lattices with gain and loss close to the parity-time symmetric condition [11]. To the second type belong the localised modes in the gap separating two photonic bands when a local potential is added to the otherwise periodic structure. This is actually the principle of photonic crystal cavities and of Tamm modes at the surface of a photonic system [12–14]. Recently, the use of bands with non-trivial topology has allowed the implementation of localised modes at the edges and corners of photonic lattices of different dimensionalities without the need of introducing local on-site potentials [15–22]. The edge or corner modes appear as a consequence of the topology of the bulk with the great asset that their existence and optical frequency are protected from certain types of disorder.

One of the main limitations to the above mentioned designs is that the underlying interference mechanism responsible for localisation presents fundamental limits for the localisation length which is, in general, larger than a single lattice site. For instance, compact localised states in flat bands arise from interference between at least two lattice sites with non-zero amplitude and opposite phases [23,24]. Similarly, bound states in the continuum involve light intensity in several sites [10,25], and localisation by disorder requires multiple scatterers. In all cases, the localised modes are eigenmodes of the system, whose amplitude distribution is independent of the external excitation conditions. Another important limitation is that the modes are localised by design, that is, the geometry of the dielectric structure sets the location, shape and extension of the localised modes. This implies that after fabrication these properties are hardly adjustable.

In this article we demonstrate a method to engineer fully reconfigurable localised modes in photonic lattices of coupled resonators, with localisation lengths covering a number of sites as small as a single site. To reach this ultimate limit, we take advantage of the interplay of coherent drive and dissipation in a lattice of lossy resonators, a mechanism that had not been previously considered to engineer localised modes. Indeed, in dissipative lattices driven at resonance, the amplitude and phase of the external driving laser provide additional knobs to engineer interference effects. By choosing the appropriate arrangement of the external drives it is possible to implement localised modes at photon energies in which the passive system (without coherent drive) would present extended modes. Recently, the interplay of drive and losses has been used to demonstrate the trapping of light in a single site of a two coupled pillars system [26]. Here, we demonstrate that the localised response of a driven-dissipative photonic lattice can be engineered with virtually any geometry. The concept is reminiscent of wavefront shaping techniques employed to obtain a desired output through a disordered medium [27–29]. The localised modes demonstrated here in a photonic lattice can be seen as fully reconfigurable optical cavities defined by properly designing the external control field. We experimentally illustrate our method using a honeycomb lattice of coupled micropillars resonantly driven by several laser spots at energies within the photonic bands of the structure, resulting in the localisation of light on a single site. Our results can be extended to a variety of mode configurations and to almost any lattice geometry in one, two and three dimensions. They open unprecedented perspectives for the manipulation of light in photonic structures, in particular for enhancing local light-matter interactions and nonlinear effects with single site precision.

**Localisation by drive and dissipation**

To demonstrate the principle of localisation by drive and dissipation, we consider a lattice of coupled photonic resonators. Each of them is subject to radiative losses to the environment and can be driven by an external laser (coherent field). An archetypical system implementing this situation is a lattice of coupled semiconductor micropillars [30–33], which is the platform we use for the experimental realisation. The dynamics of the photon field in a lattice of coupled micropillars in the tight-binding limit can be described by the following set of coupled equations [34]:

$$i\hbar \frac{\partial \psi_m}{\partial t} = \varepsilon_m \psi_m + \sum_n t_{m,n} \psi_n - i \frac{\hbar}{2\tau} \psi_m + F_m e^{-i\omega_p t}. \quad (1)$$

$\psi_m$ is the field amplitude at the centre of micropillar $m$, $\varepsilon_m = \varepsilon_0$ is the energy of the considered mode in each pillar – assumed to be identical for all sites –, $t_{m,n}$ is the coupling amplitude between different sites of the lattice, $\tau$ is the radiative photon lifetime in each micropillar, and $F_m$ is the complex amplitude of the resonant excitation laser at site $m$ with photon energy $\hbar\omega_p$.

Equation (1) has a family of localised solutions for specific configurations of the drive field $F_m$ in the steady state. Figure 1a shows the simplest example of a one-dimensional lattice of coupled resonators with nearest-neighbours hopping amplitude $t_{m,m+1} = t_{m,m-1} = t$ and $\tau = 10\hbar/t$. If a single site in the middle of the lattice is pumped by a laser at any frequency within the photonic band, the steady state intensity simulated using Eq. (1) extends over many lattice sites (on the order of $t\tau/\hbar$ for the particular case of $\hbar\omega_p = \varepsilon_0$). Remarkably, in the configuration of Fig. 1b, in which two sites $M - 1$ and $M + 1$ are pumped with equal phase and amplitude at $\hbar\omega_p = \varepsilon_0$, the response of the lattice is almost fully localised at site $M$, which is surrounded by the two pumps. The field intensity in the pumped sites is almost zero, as well as in all lattice sites located out of the region defined by the two pumps (it tends strictly to zero for long lifetimes when $\hbar/\tau \ll t_{m,n}$). This behaviour is reminiscent of Fabry-Perot bound states in the continuum based on coupled resonances [8,10,25].

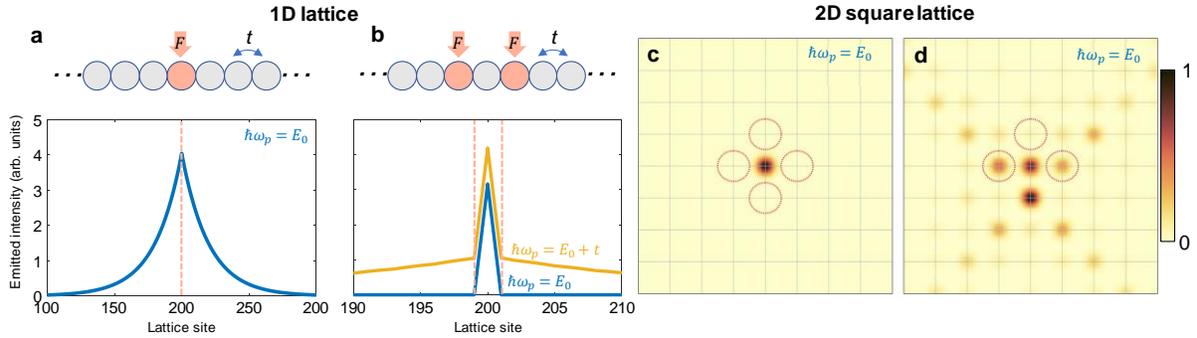

**Figure 1. Localised modes in a driven-dissipative lattice. a** One-dimensional lattice with 400 sites resonantly driven by a laser located at a single site at the energy of the middle of the photonic band, showing propagation of the emitted intensity with a decay length of $\sim t\tau/\hbar$ sites, with $\tau = 10\hbar/t$. **b** When two drives of same amplitude and phase envelope a single site, the emission is fully localized in the middle site (blue line). Away from the resonance frequency $\hbar\omega_p = \varepsilon_0$, emission extends over the whole lattice (orange line). Vertical lines mark the position of the pump lasers. **c** Calculated emission intensity in a squared lattice of 30x30 sites (zoomed view) driven at the four sites marked in red with a frequency $\hbar\omega_p = \varepsilon_0$ and identical phases. The lattice sites are located at the line crossings. **d** When one of the drive lasers is removed, emission spreads over the lattice.

This kind of localised mode is a general feature of Eq. (1): when a region of the one-dimensional lattice is delimited by two pump spots of same absolute amplitude $|F|$, it can be shown that in the limit $\tau \gg t$ there exists a discrete set of photon frequencies of the driving field for which the photon amplitude is exactly zero at the pumped sites, and different from zero within the region delimited by the pumps (see Supplementary information). Because the pumped sites have zero amplitude, the pump frequency at which this phenomenon takes place coincides precisely with the eigenenergies of the region confined by the pumps as if it was fully detached from the lattice. In the case of Fig. 1b, this is the eigenenergy of a single site decoupled from the rest of the lattice: $\hbar \omega_p = \varepsilon_0$. Simultaneously, the zero field amplitude in the pumped sites results, strictly, in zero amplitude of the photon field outside the confined region (see Supplementary information for an analytic proof). This means that the presence of the localised mode is independent of the lattice size out of the confined region. Note that if the frequency of the drive is shifted from the resonant condition, the emitted intensity is distributed over all lattice sites as attested by the slow decay of the emission out of the central peak depicted by the orange line in Fig. 1b.

The description we have just presented and the corresponding simulations of Eq. (1) can be directly extended to higher dimensional lattices. Figure 1c displays a simulation of a square lattice with four pumped sites (marked with red circles) of equal amplitude and phase surrounding a single site. A extreme localised response at $\hbar \omega_p = \varepsilon_0$, down to a single site is observed, with almost zero amplitude in the pumped sites and in the sites away from the pumped region. As a counterexample, Fig. 1d shows the case when the pump spots do not fully surround a central site. In this situation the real-space distribution of the field does not show any confined response at $\hbar \omega_p = \varepsilon_0$ or at any other laser energy.

**Experimental realisation**

Semiconductor micropillars are an ideal system to probe the concept we have just introduced of localisation by drive and dissipation. Single micropillars can be laterally etched from planar microcavities made of two AlGaAs Bragg mirrors embedding a GaAs cavity spacer. In addition, in our structures, a single InGaAs quantum well is grown at the centre of the cavity. At 6K, the temperature of the experiments, quantum well excitons are strongly coupled to the lowest optical mode of the micropillar giving rise to exciton-polaritons [34]. However, we work at sufficiently negative photon-exciton detuning for polaritons to have a 99% photonic content, and all the effects we report here can in principle be observed in structures without excitonic resonances. Each micropillar confines discrete photonic modes which can be laterally coupled in the form of a lattice by making the micropillars overlap. Here we consider a honeycomb lattice of micropillars of 2.75 μm in diameter and a centre-to-centre separation of 2.3 μm (see sketch in Fig. 2a). Angle resolved photoluminescence with out of resonance laser excitation at 1.535 eV in a 2.3 μm spot (full width at half maximum) centred on top of a micropillar reveals the two lowest energy bands of the structure (Fig. 2b). They display two Dirac crossings characteristic of honeycomb lattices [30]. The experiments are realized in transmission geometry employing linearly polarized excitation and parallel detection. No significant polarisation splitting is observed in the studied photonic bands. We fit the measured bands to a tight binding model [35] with nearest-neighbour hopping *t* = 328 μeV and next-nearest neighbour hopping *t'* = -42 μeV. The emission energy at the Dirac crossings is $E_0 = 1.3917$ eV.

Under resonant laser excitation the dynamics of the photon field in the lattice can be accurately described by Eq. (1). This description captures the driven-dissipative nature of the micropillar system, and differs significantly from the description of the dynamics of light in lattices of coupled waveguides or ring resonators, which are weakly coupled to the environment. The latter are described by a conservative version of Eq. (1) and miss the pump and loss terms. Other experimental systems such as mechanical resonators are well described by Eq. (1), but they have been mostly studied in a different context of nonlinear effects with drive extended to all lattice sites [36].

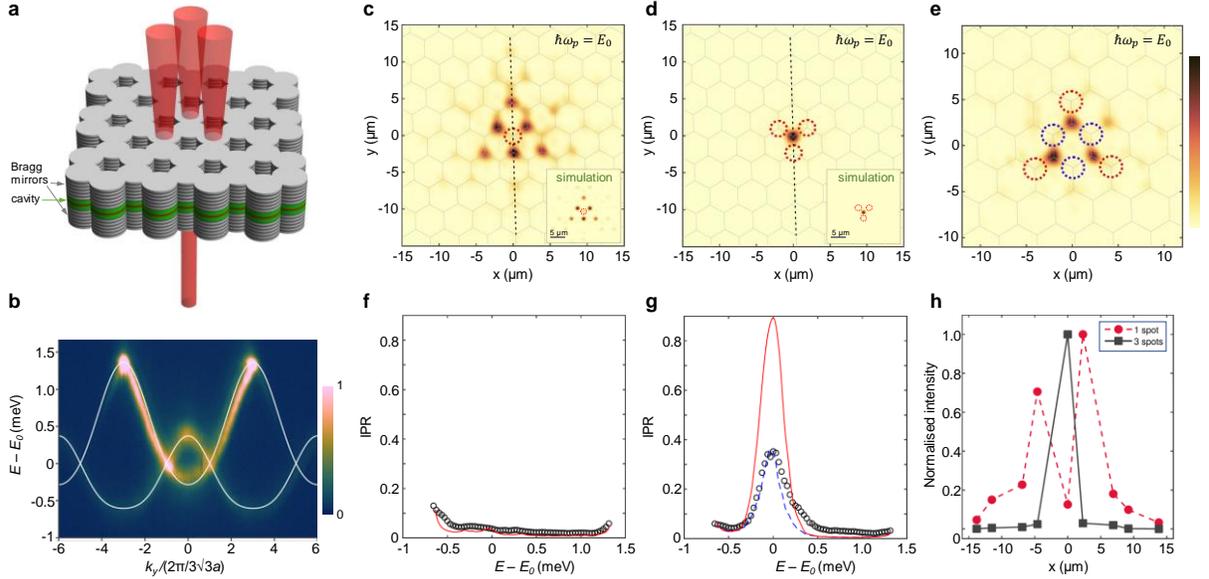

**Figure 2. Resonant drive of a honeycomb lattice. a** Scheme of the honeycomb lattice of micropillars and of the excitation (top) and emission (bottom) beams. **b** Measured angle-resolved photoluminescence of the lattice showing the energy bands as a function of in-plane momentum $k_y$ for $k_x = 2\pi/3a$, with $a$=2.3 μm the centre-to-centre separation between adjacent micropillars and $E_0 = 1.3917$ eV. White lines represent the tight-binding fit. **c** Measured emission pattern when driving the pillar marked with a red circle with a laser at $\hbar\omega_p = E_0$. The centre of each micropillar is located at the vertices of the grey hexagonal pattern. **d** Same when driving three pillars with equal amplitude and phase surrounding a central one, which shows high emission intensity. Insets in **c-d** show simulations in the conditions of the experiment. **e** Optical response when arranging three times the localization building block shown in **d** to form a staggered triangle. Laser spots in blue sites have twice the intensity of spots in the red sites. **f-g** Measured IPR (dots) as a function of the laser frequency with the pump spot configurations of **c-d**, respectively. Red lines show the calculated value using the photon lifetime of the cavity. Blue dashed line in **g** accounts for a phase difference of $0.09\pi$ between the bottom pump spot and the left one and between the right one and the bottom one. **h** Measured intensity profiles along the dashed lines in **c** and **d**.

Figure 2c displays the measured intensity when pumping a single site (red circle) of the honeycomb lattice with a laser at a frequency $\hbar\omega_p = E_0$, the energy of the Dirac crossing in Fig. 2b. The field amplitude extends away from the pump spot in a triangular shape mostly in the sites belonging to the sublattice opposite to that of the pumped micropillar. Remarkably, the pumped site shows almost no intensity. This peculiar response is a consequence of the interference effect at the pumped site between the pump laser and the eigenmodes of the lower bonding band and the upper antibonding band. This effect was already noticed in the simpler case of two coupled micropillars pumped at an energy right in between the bonding and anti-bonding modes [26], and also in a honeycomb lattice under strain [37]. The field distribution is very different when three laser spots of same amplitude and phase drive the lattice in a triangular geometry surrounding a single site. This situation is reported in Fig. 2d using a spatial light modulator to generate the pump spots. Similar to the simulations for the square lattice shown in Fig. 1c, the field distribution is strongly localised in the micropillar surrounded by the three pump spots and no significant emission anywhere else in the lattice is observed (including the pumped sites), see Fig. 2h. The measured real space patterns are well reproduced by numerically solving Eq. (1) in the steady state regime with a polariton lifetime $\tau = 9$ ps, see insets of Figs. 2c and 2d.

The extreme localised response, down to a single micropillar, can be quantified by computing the inverse participation ratio IPR= $\sum_m |\psi_m|^4 / \sum_m |\psi_m|^2$, which has a value of 1 for emission fully localised in a single site and 0 for extended modes in an infinite lattice. Figure 2g depicts the measured IPR for the three laser spots excitation as a function of the laser photon energy from the bottom to the top of the Dirac bands. The IPR is computed from the emission measured at the centre of each micropillar. The transmitted pattern is highly localised at the energy of the Dirac point ($E_0$; IPR=0.35). This photon energy is very close to the estimated eigenenergy $\varepsilon_0$ of the fundamental mode of a single detached micropillar ($E_0 = \varepsilon_0 + 3t'$), and they would both coincide exactly in the absence of next-nearest-neighbour coupling. Simulations of Eq. (1) are displayed in red lines and predict a value of IPR=0.89 for the measured polariton lifetime. The difference with the measured value of 0.35 arises from an unintentional horizontal tilt of the incident laser beams, which induces an estimated phase difference of about $0.09\pi$ between the three consecutive spots. In the limit of negligible losses, simulations show that the IPR grows asymptotically towards 0.9 (it would be 1 in the absence of next-nearest-neighbours hopping, see Supplementary information). Moreover, localisation is highly preserved in presence of mild disorder (see Supplementary information). In contrast, Fig. 2f shows that for a single spot excitation, the transmitted signal is extended over several lattice sites for any photon energy resulting in very low IPR values.

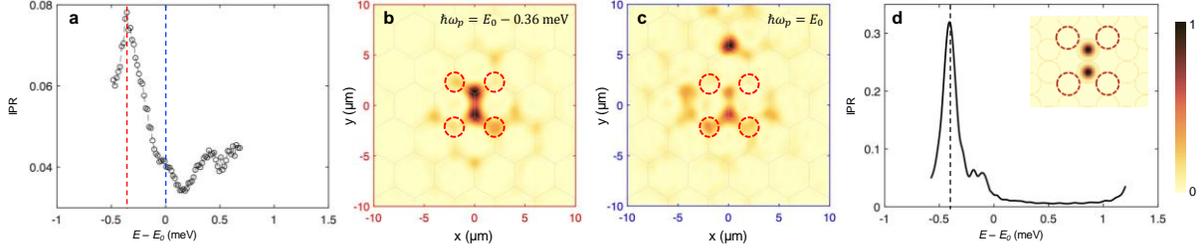

**Figure 3. Localised bonding mode. a** Measured IPR when four driving spots of nominally the same amplitude and phase demarcate a two-site molecule. For photon energies above $E_0 + 0.7$ meV residual scattered light prevent a confident measurement of the IPR. **b** Measured emitted intensity at the IPR peak (red dashed line in **a**): $\hbar\omega_p = E_0 - 0.36$ meV. The pump spots are drawn in red circles. **c** Measured emission intensity at $\hbar\omega_p = E_0$ (blue dashed line in **a**). **d** Simulated IPR in the case of perfect optical alignment. The inset shows the simulated intensity distribution at $\hbar\omega_p = E_0 - 0.407$ meV (corresponding to the dashed line in the main panel).

The localised mode with three pumping spots shown in Fig. 2d can now be used as a building block to engineer any possible intensity pattern in the lattice, just by adding groups of three laser spots of the same phase at the energy of the Dirac points. An example is shown in Fig. 2e with three groups of three excitation spots each, resulting in a staggered triangle.

So far we have discussed single site localised modes, which appear close to the energy of an isolated micropillar. If a bigger drive-induced cavity is considered, the resonant localised modes appear at the molecular eigenenergies of the cavity. To explore this situation, we now move to a configuration of pump spots of equal amplitude and phase surrounding completely two adjacent pillars of the honeycomb lattice, as sketched in red circles in Fig. 3b. In that configuration we would expect a localised mode at the energy of the bonding states of two coupled isolated sites $\hbar\omega_p = E_0 - t$. Figure 3a displays the measured IPR as a function of the laser energy from bottom to top of the photonic bands. A peak of localisation appears at $\hbar\omega_p = E_0 - 0.36$ meV, which is very close to the expected value (the difference arising from the presence of next-nearest-neighbour hopping). Despite the apparently low value of the measured IPR, caused by a slight misalignment of the excitation spots, localisation in the bonding-like mode is remarkable, as shown in Fig. 3b. As a comparison, Fig. 3c shows the measured spatial pattern at $\hbar\omega_p = E_0$, displaying a significant spread. Under perfect alignment conditions, the IPR value should go up to 0.32 (see Fig. 3d). Localisation at the energy of the antibonding two-sites mode of the confined molecule is observed at $\hbar\omega_p \approx E_0 + t$ when the upper two pump spots have a phase difference of $\pi$ with respect to the two lower pump spots (see Supplementary information).

As demonstrated above, the photon energy at which the localised mode takes place is determined by the eigenenergies of the optically confined cavities as if they were detached from the lattice. We will now show that the resonance energy for high localisation can be modified at will, at least in the simplest cases, by adding an additional pump spot on top of the localised sites. We consider again the situation with three identical pump spots depicted in Fig. 2d, which results in localisation at a single site. On top of this localised site we add an additional drive of the same frequency as the surrounding laser spots. Remarkably, the resonance frequency for the localisation is now modified as shown in simulations in Fig. 4a. The magnitude of the frequency shift is determined by the relative amplitude of the added spot $F_{in}$ compared to that of the surrounding spots $F_{out}$. The sign of the shift, towards higher or lower energies, is set by its phase, 0 or $\pi$ with respect to the phase of the surrounding pump spots. For the photon lifetime of our experiments, the IPR of the localised mode has some variations with frequency related to the shape of the photon dispersion and the density of states of the lattice. Figure 4b shows that these variations are smoothed for long photon lifetimes.

Figure 4c shows the measured IPR as a function of laser frequency when the intensity of the additional central spot is three times larger than that of the surrounding spots and, nominally, the same phase. The IPR peak is shifted from $E_0$ to $E_0 + 0.69$ meV. Figure 4d shows highly localised emission at a single site at that photon energy.

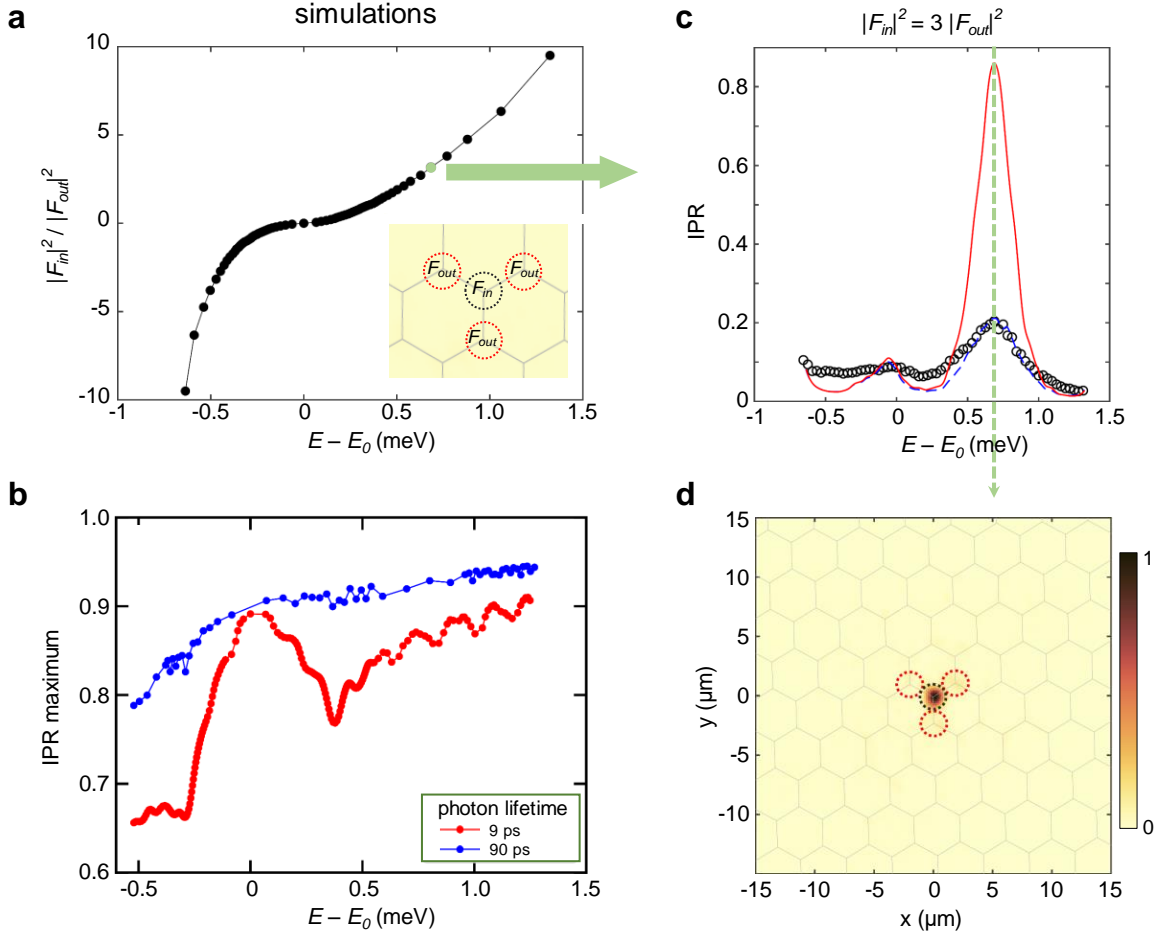

**Figure 4. Modifying the energy of the localised mode. a** Calculated photon energy of the IPR peak for the four spots configuration shown in the inset. The vertical axis shows the ratio of intensities of the central pump spot $|F_{in}|^2$ to the outer three spots $|F_{out}|^2$. The sign represents the relative phase between outer and inner spots: positive means same phase, negative phase means a phase difference of π. **b** Calculated value of the IPR maximum when its energy is peaked at the value indicated in the horizontal axis for two values of the photon lifetime in the resonators. **c** Measured IPR (dots) when scanning the laser frequency in the configuration of the inset in **a** for $|F_{in}|^2 = 3 \times |F_{out}|^2$ (dashed line in **a**). The red line is the calculated IPR using the photon lifetime of the lattice and equal phase for the four spots. The blue line includes a phase difference of $0.05\pi$ for one of the outer spots. **d** Real space emission measured at the energy of the IPR peak ($E - E_0 = 0.69$ meV).

**Discussion**

We have demonstrated that the combination of resonant drive and dissipation in lattices of coupled photonic resonators can be advantageously used to design highly localised emission patterns in a reconfigurable manner with high flexibility. Alternatively to the comprehensive design scheme we have presented here, reverse engineering can be directly employed by solving Eq. (1) in the steady regime for the excitation fields $F_m$ after imposing a desired shape and frequency of the photon field in the lattice. These features can be directly transposed to other photonic systems such as lattices of superconducting microwave cavities, resonators with spectral synthetic dimensions [38], (opto)mechanical lattices and, more generally, to fluid wave systems. They could be used to locally enhance nonlinear effects, for instance, in polariton lattices with high exciton content, and for the control of light-matter interactions that require single site excitation in a dense matrix. This is the case, for instance, when the resonator cavity contains single photon emitters whose emission properties may vary from site to site [39].

Interestingly, the kind of localised modes observed here are analogous to the localised emission patterns expected from quantum emitters in photonic lattices discussed in Refs. [40–42]. In our realisation, the role of the quantum emitter coupled to the lattice is played by the resonant pump spots. Beyond localised modes, those theoretical works have shown the possibility of engineering highly directional responses in the lattice and that the decay dynamics can be strongly modified by the photonic density of states. These promising ideas could be directly transposed to the configuration discussed in our work in a purely photonic realisation and enlarge the possibilities of manipulating light-matter interactions in lattices of resonators.

## Methods

**Sample description.** The microcavity used for these studies was fabricated by molecular beam epitaxy. The heterostructure is made of two Bragg mirrors of 28 (top) and 32 (bottom) pairs of alternating layers of $Ga_{0.90}Al_{0.10}As$ and $Ga_{0.05}Al_{0.95}As$ of thickness $\lambda/4$, embedding a $\lambda$ spacer of GaAs, with $\lambda = 880$ nm/$n$, and $n$ being the index of refraction of the material of each layer at 6K. A single $In_{0.09}Al_{0.91}As$ quantum well of 20 nm in width is grown at the centre of the cavity spacer. The whole heterostructure is grown on an epitaxial quality GaAs substrate. During the growth of the Bragg mirrors and the cavity spacer the substrate was not rotated. Due to the flux angle of the different material cells in the growth chamber with respect to the substrate, the cavity and mirrors present a wedge across the wafer. In this way, different points of the wafer have different thicknesses and different cavity-mode energies. In the reported experiments we select a point of the wafer in which the lowest photonic modes is red-detuned by 18 meV from the quantum well exciton resonance (at 1.4099meV at 6K, the temperature of our experiments). This detuning is much larger than half the Rabi splitting 3.5meV/2=1.75meV. In this way, any polaritonic effect can be neglected in the present experiments and all our data imply bare photons. By selecting a different point of the sample with a smaller photon-exciton detuning, it would be possible to work with polaritons with higher excitonic content and address, for instance, nonlinear effects.

The as-grown planar structure is then processed with electron beam lithography and it is etched down to the substrate (Inductively Coupled Plasma Etching). By properly designing the electron beam mask, we fabricate honeycomb lattices of coupled round micropillars with 2.75 μm diameter and a centre-to-centre separation of 2.3 μm. In the lattices considered in this work, the photon lifetime is estimated to be 9 ps. This value is obtained from fitting simulations of Eq. (1) in the conditions of the experiment to the observed propagation patterns in Fig. 2c.

**Experimental setup.** The sample with the lattice of micropillars is held in vacuum inside a closed-cycle He cryostat at a temperature of 6K as measured on the sample holder. The excitation beam is a Ti:Sapph single mode laser (<10MHz linewidth) which passes through a single-mode polarisation maintaining fibre to obtain a clean Gaussian mode. After the fibre, the beam goes through a spatial light modulator in reflexion geometry, a telescope and the excitation lens, which allow engineering the excitation pattern in the form of one, three or more spots on the surface of the lattice. The experiment is done in transmission geometry: the excitation beam impinges the sample through the epitaxial side, traverses the substrate, passes through a slit of 3 mm width in the sample holder and exits the cryostat towards the collection optics. Both for excitation and detection we use 8 mm focal length aspherical lenses with numerical aperture 0.5. Note that the GaAs substrate is transparent with negligible absorption at the wavelength of this work.

After being collected by the collection lens, the emitted light is imaged onto a CCD and a spectrometer. The excitation is vertically polarised and detection is done along the same polarisation axis.

**Simulations.** The numerical simulations of Eq. (1) displayed throughout the article have been realized using the split-step method after switching on the excitation field in a step function. We wait long enough to obtain the steady state response. To display the result of the simulations in space, each point of the lattice shows the amplitude square of the fundamental Laguerre-Gauss mode with a Full Width at Half Maximum of 2.4 μm, with a peak intensity given by the simulation.

**Acknowledgements**

We thank Fabrice Lemoult for fruitful discussions. This work was supported by European Research Council grants EmergenTopo (865151) and ARQADIA (949730), the H2020-FETFLAG project PhoQus (820392), the QUANTERA project Interpol (ANR-QUAN-0003-05), the Paris Ile-de-France Région in the framework of DIM SIRTEQ, the French National Research Agency project Quantum Fluids of Light (ANR-16-CE30-0021), the French RENATECH network, the French government through the Programme Investissement d'Avenir (I-SITE ULNE / ANR-16-IDEX-0004 ULNE) managed by the Agence Nationale de la Recherche, the Labex CEMPI (ANR-11-LABX-0007) and the CPER Photonics for Society P4S. K.S. acknowledges the doctoral scholarship ETIUDA (No. DEC-2019/32/T/ST3/00332) financed by the Polish National Science Centre (NCN). A.G.-T. acknowledges financial support from the Proyecto Sinérgico CAM 2020 Y2020/TCS-6545 (NanoQuCo-CM), the CSIC Research Platform on Quantum Technologies PTI-001 and from Spanish project PGC2018-094792-B-100(MCIU/AEI/FEDER, EU).

**Competing interests**

The authors declare no competing interests

# Supplementary information

**Theoretical analysis of localisation by drive and dissipation in a 1D lattice**

To get an intuitive understanding of the nature of the localisation of the emission in the polariton lattices subject to drive and dissipation, we analyse the driven-dissipative Schrödinger Eq. (1) in the main text:

$$i\hbar \frac{\partial \psi_m}{\partial t} = \varepsilon_m \psi_m + \sum_n t_{m,n} \psi_n - i\frac{\hbar}{\tau}\psi_m + F_m e^{-i\omega_p t}. \qquad (S1)$$

$\psi_m$ is the field amplitude at the centre of micropillar $m$, $\varepsilon_m = E_0$ is the energy of the lowest energy mode in each pillar –assumed to be identical for all sites–, $t_{m,n}$ is the coupling amplitude between different sites of the lattice, $\tau$ is the photon radiative lifetime, and $F_m$ is the complex amplitude of the resonant excitation laser at site $m$ with photon energy $\hbar\omega_p$.

In the rotating frame of the pump frequency $\omega_p$, in the limit of losses much weaker than the hopping (i.e., $\frac{\hbar}{\tau} \ll t_{m,n}$), the term $-i\frac{\hbar}{\tau}\psi_m$ can be neglected and the steady state of Eq. (1) takes the simple form:

$$\Delta \psi_m + \sum_n t_{m,n} \psi_n = -F_m, \qquad (S2)$$

where $\Delta = E_0 - \hbar\omega_p$ is the detuning between the onsite energy and the pump frequency. We consider the case of a one-dimensional lattice with nearest-neighbour hoppings in the configuration described in Fig. 1(b), with pump at lattice sites $M - 1$ and $M + 1$ with drive of amplitudes $F_{M+1}$ and $F_{M-1}$.

We can explicitly write Eq. (S2) for a few sites at an around the pumps:

| | | |
|---|---|---|
| site $M$: | $\Delta\psi_M + t(\psi_{M-1} + \psi_{M+1}) = 0$ | (S3) |
| site $M - 1$: | $\Delta\psi_{M-1} + t(\psi_{M-2} + \psi_M) = -F_{M-1}$ | (S4) |
| site $M + 1$: | $\Delta\psi_{M+1} + t(\psi_M + \psi_{M+2}) = -F_{M+1}$ | (S5) |
| site $M + 2$: | $\Delta\psi_{M+2} + t(\psi_{M+1} + \psi_{M+3}) = 0$ | (S6) |
| site $M + 3$: | $\Delta\psi_{M+3} + t(\psi_{M+2} + \psi_{M+4}) = 0$ | (S7) |
| site $M + 4$: | $\Delta\psi_{M+4} + t(\psi_{M+3} + \psi_{M+5}) = 0$ | (S8) |
| site $M + 5$: | $\Delta\psi_{M+5} + t(\psi_{M+4} + \psi_{M+6}) = 0$ | (S9) |

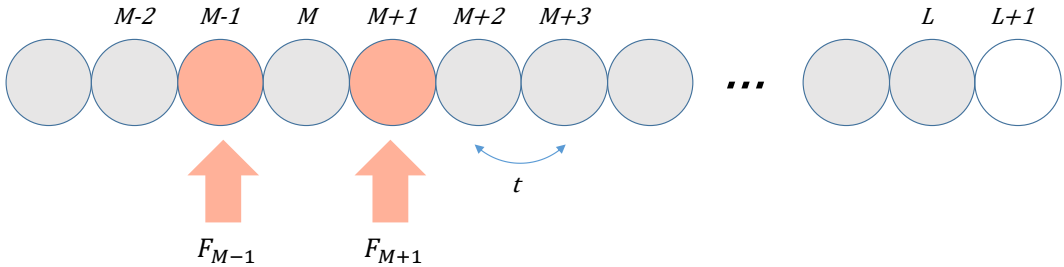

**Fig. S1**. Schematic representation of a one-dimensional lattice with two pump drives at sites $M - 1$ and $M + 1$. The site $L + 1$ is out of the lattice.

Let us look for solutions for which the pumped sites $M \pm 1$ have zero amplitude ($\psi_{M-1} = \psi_{M+1} = 0$). From Eq. (S3) we see that this implies that $\Delta = 0$. From Eq. (S6) we see that if $\Delta = 0$ and $\psi_{M+1} = 0$, then $\psi_{M+3}$ must be zero. Subsequently, Eq. (S8) implies that $\psi_{M+5} = 0$ and this happens for all the sites $m = M \pm (2n+1)$, for $n = 1, 2, 3, ...$ Therefore, all sites separated by an odd number of pillars from the central site $M$ must have zero amplitude.

We will see now that this is the case also for sites separated by an even number of pillars from the central site $M$. From Eq. (S7) and (S9) with $\Delta = 0$, we see that $\psi_{M \pm 2n} = \psi_{M \pm (2n+2)}$. Consequently, all these sites have the same amplitude. We can see that their amplitude must be zero using the following argument: let us assume that the last site of the lattice, $L$, is separated by an odd number of pillars. Then, the site $L+1$, which actually does not exist, can be described by imposing its amplitude to be zero[1]: $\psi_{L+1} = 0$. Therefore, all the series of sites $\psi_{M \pm 2n}$ to which it belongs are also zero.

To calculate the amplitude in the central site $M$ we focus on Eqs. (S4) and (S5). As we have just seen, if $\Delta = 0$ there is a solution to the set of coupled Eqs. (S2) in which all sites except for $\psi_M$ are zero. Inserting $\psi_{M+1} = \psi_{M-1} = 0$ in Eqs. (S4) and (S5) we directly see that $\psi_M = -F_{M-1}/t = -F_{M+1}/t$: the pump spots on the two sites surrounding $M$ must have equal amplitude and phase for the destructive interference effect away from $M$ to take place. The amplitude at $\psi_M$ is directly proportional to the pump amplitude.

**Multiple sites**

The interference of the pump fields in the lattice has a double effect. First, from the previous example, we see that the search for solutions in which the pumped sites have zero amplitude implies that the frequency of the pump must coincide with the eigenenergies of the site surrounded by the pumped sites, as if it was detached from the lattice. Second, the condition that the pumped sites have zero amplitude implies that the amplitude is zero everywhere else in the lattice except from the sites surrounded by the pumps.

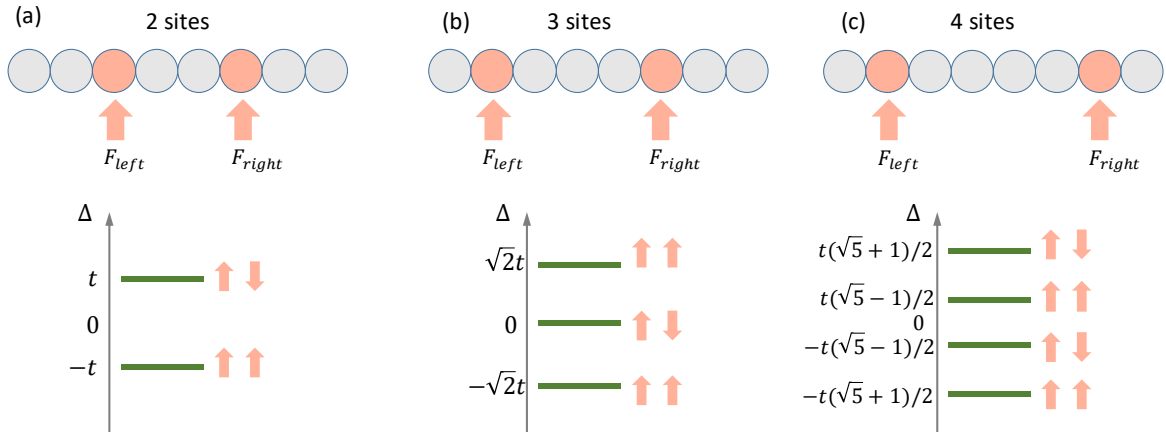

**Fig. S2**. Scheme for the generation of localised modes in larger chain sections, when the driving pumps are separated by 2, 3, and 4 sites in, (a), (b), and (c), respectively. The lower panels show the resonant driving frequencies of the localised modes. They correspond to the eigenenergies of 2, 3 and 4 coupled sites, respectively. The arrows indicate the relative phase of the pumps to excite each respective mode (up-up are in-phase pumps, up-down are out-or-phase pumps).

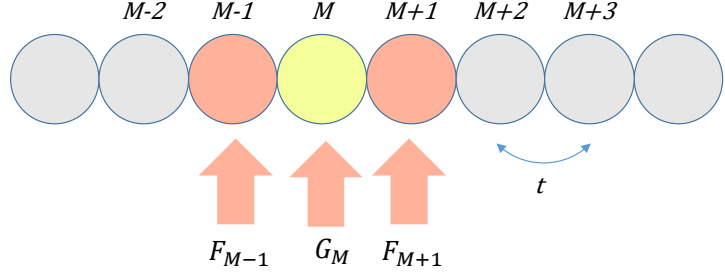

**Fig. S3.** Schematic representation of a one-dimensional lattice with two pump drives at sites $M-1$ and $M+1$ and an additional drive at site $M$.

We can extend this argument to situations in which the pumps are separated by more than two sites. Figure S2 shows several cases along with the pump energies at which modes entirely localised in between the pump spots are found. When the region in-between pumps is made of two sites (Fig. S2(a)), the localised modes appear at frequencies $\Delta$ corresponding to the eigenmodes of two coupled isolated sites: $\Delta = -t$ and $\Delta = +t$. The amplitude and phase of the two pumps must respect the phase distribution of the isolated modes at the edges: for $\Delta = -t$ (bonging mode) $F_{left} = F_{right}$, for $\Delta = t$ (antibonging mode) $F_{left} = -F_{right}$. Note that this situation is very similar to the case reported in Fig. 3 of the main text for two pillars in a honeycomb lattice.

If the region in-between pumps is larger, more localised resonances appear (see Fig. S2(b)-(c)), with frequencies given by the eignemodes of the region in-between the pumps as if it was detached from the lattice.

**Shift of the energy of the localisation resonance**

Figure 4 in the main text shows that the drive frequency of the localisation resonance at a single site can be shifted at will by adding an extra drive on top of the localisation site. In the main text we have proved this feature numerically for the conditions of Fig. 4. To get further insights, we now consider the one-dimensional lattice of Fig. S3 and Eqs. (S3)-(S9). We treat the case in which $F_{M-1} = F_{M+1} = F$ and we add an additional pump of same photon energy $\hbar\omega_p$ and amplitude $G_M$ to the site surrounded by the two pumps $F$. As mentioned in the main text, we anticipate that this situation will modify the photon energy at which the perfect localisation condition takes place. In the present case, all equations (S3) to (S9) remain the same except for Eq. (S3), which becomes:

$$\text{site } M: \qquad \Delta\psi_M + t(\psi_{M-1} + \psi_{M+1}) = -G_M \qquad (S3')$$

If we now impose the condition $\psi_{M-1} = \psi_{M+1} = 0$, we see that $\psi_M = -G_M/\Delta$. Inserting this into Eq. (S5) gives:

$$G_M = \frac{F_{M+1}\Delta}{t} + \psi_{M+2}. \qquad (S10)$$

We now look for the condition that $\psi_{M+2}$, which is out of the region in-between the pumped spots, is zero. This imposes a value of the drive detuning:

$$\Delta = t\frac{G_M}{F_{M+1}} = t\frac{G_M}{F} \tag{S11}$$

For that value of the drive detuning with respect to the energy of an isolated micropillar, the fields at the sites under the external pumps $M \pm 1$ and at site $M + 2$ are zero. The same argument using Eq. (S4) shows that the field is also zero at $M + 2$. Now, using the recurrent equations (S6), (S7), (S8), … we can see that, for the similar reasons as in the previous example, the amplitude $\psi_m$ all sites with $m > M + 1$ and $m < M - 1$ must be zero.

Equation (S11) thus shows that the photon energy $\hbar\omega_p = \Delta + E_0$ of the driving field at which localisation takes place in a single site can be modified by adding an extra laser of amplitude $G_M$ at the site surrounded by the main driving fields. If the extra laser is in phase with the pumps $F$ ($G_M > 0$), the resonance condition moves to higher $\Delta$, that is, to higher laser frequency energies. While an out of phase field ($G_M < 0$) shifts the resonance condition to lower energies. The additional field $G_M$ acts as a renormalized onsite energy for the isolated site.

**Extension to 2D lattices**

The previous discussion aims at providing intuitive insights on the interference process at the core of the localisation phenomena we report. We have treated the simple case of a one-dimensional lattice with simple pump configurations. We anticipate that these results can be extrapolated to other pump configurations and to higher dimensional lattices. Several examples including a square lattice and a honeycomb lattice are treated in the main text.

In the present work we have considered situations in which the pump spots surround the region in which localisation resonances takes place. It would be interesting to explore other situations that go beyond this geometry. For instance, recent theoretical studies of the behaviour of a quantum emitter in a photonic lattice show that just a few number of emitters cleverly placed can confine light in large areas of the lattice[2–6]. Even a single emitter close to a corner of a two-dimensional lattice has been shown to do this[5]. The similarity between the two systems (the role of the emitters is here played by the resonant drives) anticipates interesting perspectives for our all-optical configuration.

**Role of lifetime**

The simple analytical model we have developed above for a one-dimensional lattice assumed that photon the photon loss rate was much smaller than the hopping amplitude. This situation results in perfect destructive interference effects at and away from the pumped spots. If photon losses are significant, the interference from the fields injected by the different pumps is not perfect and localisation is degraded. Nevertheless, the robustness of the localisation phenomenon to short lifetimes is remarkable, particularly when the considered geometry involves spots that are not too far apart. Figure S4 shows, for the geometry of three pumped sites of Fig. 2d of the main text, the calculated IPR at $\hbar\omega_p = E_0$ as a function of the photon losses $\tau$. In the limit of very large losses ($\tau \to 0$) we obtain an IPR of 0.33. This is a consequence of the fact that we chose to illuminate the sample with three spots and that the light is immediately radiated before having the time to hop to the neighbouring sites. When increasing the lifetime progressively, photons have the time to go through a first hopping event and, as the initial phase of the three pumps is the same, they interfere constructively in the centre. This situation is depicted in the left inset of Fig. S4, which shows that the population is now split over 4 pillars leading to an IPR smaller than 0.33 for small but non-zero lifetime. Increasing even more the lifetime allows the light to realize larger path loops in the lattice and destructive interference effects at the pump spots set-in. Simultaneously, the population increases dramatically at the center pillar, where the interferences are constructive. In this case the IPR reaches very high values of up to 0.9 (see right inset of Fig. S4). We can see that the limitation of the maximal IPR value observed in this configuration (solide blue line) comes from the presence of next-nearest neighbour interactions. When next-nearest neighbour hopping is suppressed (dashed blus line), the IPR asymptotically reaches the value of one at very long lifetimes.

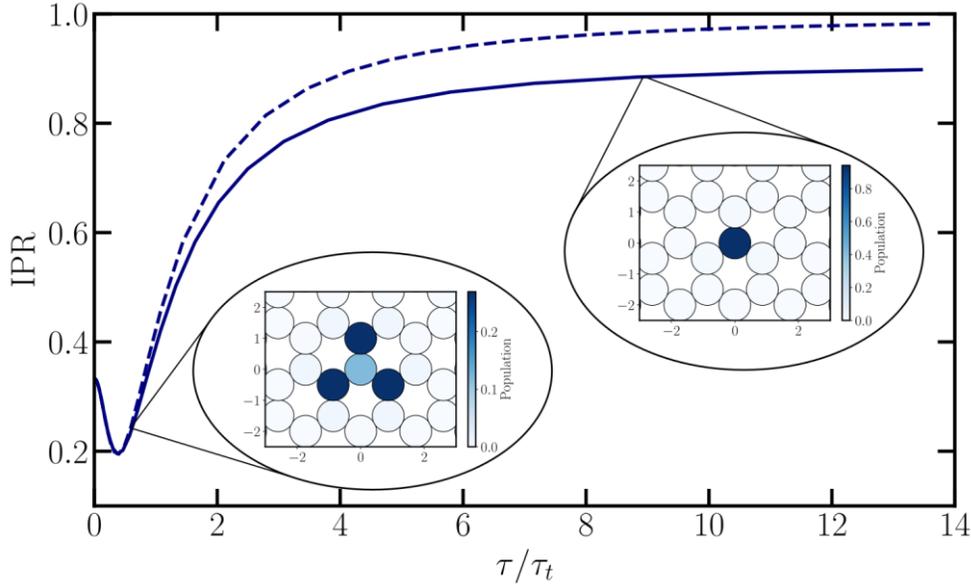

**Fig. S4.** Calculated value of the IPR peak at $\hbar\omega_p = E_0$ as a function of the photon lifetime $\tau$ in the conditions of Fig. 2(d). The dashed curve is calculated without next-nearest-neighbour coupling while the solid one takes it into account using the values of the main text: nearest-neighbour coupling $t = 0.328$ meV, next-nearest-neighbour coupling $t' = 0.042$ meV. The typical time associated to the nearest-neighbour hopping is $\tau_t = \hbar/(2t)$. The insets show the calculated populations of the real space patterns at selected values of lifetime (0.25ps and 9ps).

**Investigation of the anti-bonding scenario**

Figure S5 presents an investigation of the scenario where four pump spots of equal amplitude are used, and where the upper two spots have a phase difference of π with respect to the lower two spots. This specific configuration, where the phase is experimentally controlled with a SLM, allows to excite the anti-bonding mode of the two coupled sites molecule which is expected at a photon energy $E \approx E_0 + t$. The measured IPR (bottom panel) presents a peak at $E = E_0 + 0.221$ meV. The real space emission at that energy (top panel) clearly shows localisation in the two pillars surrounded by the four pumps. The anti-bonding character of this mode can be identified from the node in the emission at the boundary between the two pillars, which is expected if the upper and lower emit with a phase difference of π. The low value of the measured IPR peak and the presence of a peak at around $E = E_0 - 0.4$ meV (close to the bonding mode)

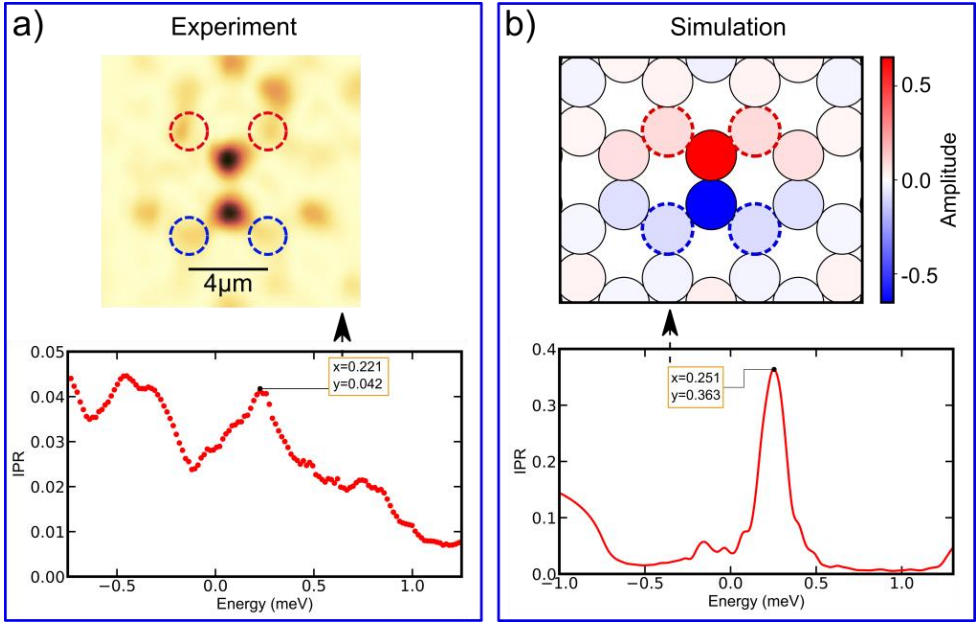

**Fig. S5.** (a) Top: Measured real space emission pattern at photon energy $E = E_0 + 0.221$ meV for excitation with four pump spots located at the sites marked with dashed lines. Nominally, upper and lower spots present a phase difference of π. Bottom: measured IPR as a function of the laser frequency. $E_0$ has been artificially shifted to zero. (b) Top: Calculated field amplitude distribution at each lattice site for the driving energy corresponding to the IPR peak at $E = E_0 + 0.251$ meV. Bottom: Calculated IPR as function of the photon energy.

suggest that the actual phases of the excitation spots depart from the nominal value.

Figure S5(b) shows the calculated IPR as a function of photon energy for perfect alignment conditions (bottom panel). An IPR peak is visible at $\hbar\omega_p = E_0 + 0.251$ meV. This value deviates from the antibonding energy of the two coupled isolated micropillars $\hbar\omega_p = E_0 + t$ (recall that $t = 0.328$ meV) due to the next-nearest-neighbour hopping included in our model ($t' = 0.042$ meV). The field amplitude at the energy corresponding to the peak of the IPR is shown in the upper panel and evidences the anti-bonding nature of the localised mode.

**Effects of disorder**

The effect of disorder on the localisation of light is shown in Fig. S6 for the case of three spot excitation. The figure shows the numerically calculated IPR in the conditions of Fig. 2d,g. The red dashed lines depict the IPR in the absence of any disorder, same phase for three spots and a photon lifetime of 9 ps. The solid lines show the calculated IPR for three different realisations of disorder in the onsite energies (a) and in the hoppings (b). In (a) the value of each onsite energy is varied randomly between $E_0 - t$ and $E_0 + t$, where $E_0$ is the energy of the central site surrounded by the three pump spots. In (b) each nearest neighbour hopping amplitude was randomly varied between $0.75t$ and $1.25t$. In both cases the IPR shows a very high peak value, attesting a high degree of localisation, and a very similar width.

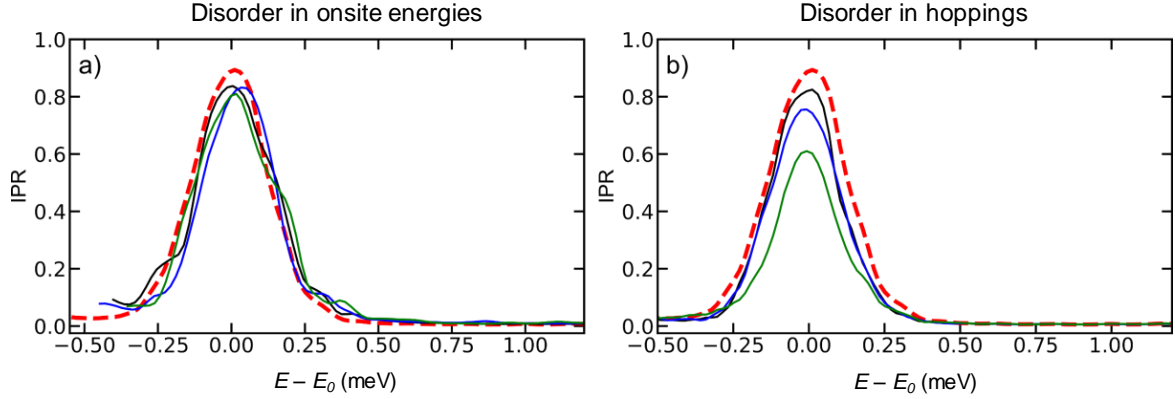

**Fig. S5.** (a) Top: Measured real space emission pattern at photon energy $E = E_0 + 0.221$ meV for excitation with four pump spots located at the sites marked with dashed lines. Nominally, upper and lower spots present a phase difference of π. Bottom: measured IPR as a function of the laser frequency. $E_0$ has been artificially shifted to zero. (b) Top: Calculated field amplitude distribution at each lattice site for the driving energy corresponding to the IPR peak at $E = E_0 + 0.251$ meV. Bottom: Calculated IPR as function of the photon energy.